\RequirePackage{lineno}
\documentclass[aps,prc,twocolumn, floatfix]{revtex4}
\usepackage{graphicx}
\usepackage{dcolumn}
\usepackage{bm}
\usepackage{multirow}

\begin{document}

%\setpagewiselinenumbers
%\modulolinenumbers[5]
%\linenumbers

\title{Strangeness production in neutron star matter}

\author{Gao-Chan Yong$^{1,2}$}
%\email[]{yonggaochan@impcas.ac.cn}

\affiliation{
$^1$Institute of Modern Physics, Chinese Academy of Sciences, Lanzhou 730000, China\\
$^2$School of Nuclear Science and Technology, University of Chinese Academy of Sciences, Beijing 100049, China
}

\begin{abstract}

Based on a dynamical model on particle production, the production and fraction of exotic components in neutron star matter are analyzed. It is found that there exists a small fraction
of strangeness in twice saturation density matter. For five times saturation density matter, the fraction of strange baryons can be as high as 25-50\%, depending on the equation of state used. The neutron-proton asymmetry of dense matter does not significantly impact the strangeness fraction in neutron star matter. This research provides new insights into the strange components in neutron stars.

\end{abstract}

\maketitle

%\section{Introduction}
%
Neutron stars are one of the densest objects in our universe, with densities exceeding that of atomic nuclei and immense gravitational fields. They are formed from the remnants of supernova explosions and are primitively supposed to be primarily made up of neutrons. However, a series of theoretical investigations have shown that neutron stars may also contain strange matter, a type of matter composed of strange quarks \cite{nk82,apj85,sch96,wz12,lon15,ch16,xia19,ger20,hu22,jj23}. This ``strangeness'' in neutron star matter has sparked interest in physical community as it could provide valuable insights into the strong nuclear force and the behavior of matter under extreme conditions \cite{wz12,xia19,jj23,sxx2023,vid18,ppnp2020}.

The presence of strangeness in neutron stars could have significant implications for our understanding of these objects \cite{nkg2001}. It could affect the internal structure and cooling of neutron stars \cite{cool92,cool99,cool3,cool4,cool5}. Strangeness could also affect the outcome of neutron star mergers \cite{mark21}, potentially producing unique gravitational wave signals \cite{gw17}. Strangeness softens equation of state of dense matter in general, thus might claim its role in the shock-wave/neutrinodelayed-shock process, presumably changing the explosion in the formation of core-collapse supernovae \cite{npa1997,soft2,soft3,soft4}. Studying the properties and abundance of strangeness in neutron stars is therefore an important area of research for astrophysicists and particle and nuclear physicists. By observing the properties and behavior of neutron stars, and comparing these with theoretical models and simulations, scientists hope to gain a better understanding of the nature of matter under extreme conditions, as well as the processes that govern the evolution of these objects over time \cite{cool99}.

There are several theoretical methods to study strange particles in neutron stars. One method involves using various nuclear many-body theories to calculate the properties of strange particles in dense matter \cite{ch16,ppnp2020}. Another method is to use effective field theories to describe the interaction between strange particles and nucleons \cite{jm14}. Additionally, some researchers use perturbative quantum chromodynamics to study strange quark matter \cite{ms1987,xiacj2017}. Also one can use astrophysical observations of neutron stars to infer the properties of strange particles within them \cite{longwh12,hell14}. In all the above methods, the interactions between strangeness and non-strangeness in dense matter are, in fact, rarely rectified by comparing with high-energy nuclear experiments in terrestrial laboratory. Relativistic heavy-ion collisions can reveal the interactions among particles at short distance or high baryonic densities. Determination of the strong nuclear force including strangeness in the context of high densities is the core question of ``hyperon puzzle'' \cite{vid18} and affects the fraction of strangeness. While the fraction of strangeness conversely affects the bulk stiffness of neutron star matter or the maximum mass of neutron stars. The interplay of the fraction of strangeness and the equation of state of neutron star matter complicates the question on ``hyperon puzzle''. To first determine the fraction of strangeness in neutron star matter, in the present study, I use a relativistic heavy-ion collision transport model, which is frequently used to simulate particle productions in heavy-ion collisions, to study the strangeness production in neutron star matter through box simulation.

%\section{The pure hadronic AMPT-HC model}
%
The used collision dynamical model is a mode of the A Multi-Phase Transport (AMPT) model \cite{AMPT2005}, which only deals with pure hadron cascade with hadronic mean-field potentials (dubbed as AMPT-HC) \cite{cas2021}. In the AMPT-HC model, the Woods-Saxon nucleon density distribution and local Thomas-Fermi approximation are used to initialize the position and momentum of each nucleon in colliding projectile and target. In addition to the usual elastic and inelastic collisions, hadron potentials with the test-particle method are applied to nucleons, baryon resonances, strangenesses as well as their antiparticles \cite{cas2021,yongrcas2022}. In the model, $\pi$, $\rho$, $\omega$, $\eta$, $K$, $K^*$, $\phi$, $N$, $\Delta$, $N^*(1440)$, $N^*(1535)$, $\Lambda$, $\Sigma$, $\Xi$ and $\Omega$ are included. Since the form of single nucleon potential at high momenta and high densities is still less known, to make minimum assumption, here we use the density-dependent single nucleon mean-field potential
$U(\rho)=\alpha\frac{\rho}{\rho_0}+\beta(\frac{\rho}{\rho_0})^\sigma$ with $\alpha=(-29.81-46.9\frac{k+44.73}{k-166.32}){~\rm MeV}$,
$\beta=23.45\frac{k+255.78}{k-166.32}{~\rm MeV}$,
$\sigma=\frac{k+44.73}{211.05}$ (where $\rho_{0}$ and $k$ stand for nuclear saturation density and incompressibility of nuclear matter, respectively) to model the stiffness of nuclear matter, i.e., the equation of state (EoS) of nuclear matter \cite{guo2021}. As for the asymmetric part of the EoS, which is crucial to neutron star matter, a form of $E_{sym}(\rho/\rho_{0})=34.5(\rho/\rho_{0})^{\gamma}$ is employed \cite{esym21}, where $\gamma$ = 0.5 or 1.5 for the default soft symmetry energy and a stiff symmetry energy as counterpart, respectively.

In the AMPT-HC model, strangeness productions from baryon-baryon,
meson-baryon as well as meson-meson scatterings were detailedly specified in Refs. \cite{AMPT2005,cas2021,yongrcas2022} and references therein. The form of kaon potential is taken from Ref.~\cite{ligq97} while no mean-field potential is used for pions. For strange baryons $\Lambda$, $\Sigma$, $\Xi$ and $\Omega$, we adopt the quark counting rule asserting that these strange baryons interact with other baryons only through their non-strange (2/3, 2/3, 1/3, 0) constituents \cite{mos74,chung2001}. Therefore, $YN$ potential is 2/3$U_{N}$ (where $Y$ stands for $\Lambda$ or $\Sigma$, $U_{N}$ is the single nucleon potential); $YY$ potential is 4/9$U_{N}$; $Y\Xi$ potential is 2/9$U_{N}$ and $\Xi\Xi$ potential is 1/9$U_{N}$. There is no mean-field potential for the three-strange-quark $\Omega$.

Since the neutron star matter is a long-standing and stable matter in the universe and one cannot simulate particle-particle scatterings relating to strangeness productions in box using infinite time steps. In practice, I employ a finite time steps which can saturate strangeness production quickly without introducing Pauli-blockings. Introducing the Pauli-blockings would prolong the saturation time in box simulations but does not evidently affects the results in the present study. This is because the Pauli-blocking just prevents a scattering with a larger probability but it would happen if it has more chances to scatter. If the reaction time is infinite, the Pauli-blocked scattering would have more chances to scatter. In the simulations, for a specific scattering channel, besides net charge conservation, the net strangeness is conserved. However, during the collisions, some strangenesses are short life and decay soon, the net strangeness is thus not conserved in the whole simulation process.

%\section{Results and Discussions}
%
Nuclear matter computation given by nuclear many-body models is frequently seen in the studies of properties of neutron star matter as well as bulk properties of nuclear matter in nucleus or heavy-ion collisions. Around nuclear saturation density, since there are a lot of empirical values from nuclear experiments to use, most nuclear many-body models could give reasonable predictions. While beyond saturation density, model extrapolations usually give divergence, simply because the strong interactions among different particles used at high baryon densities in nuclear many-body models are less confirmed.
Hadronic transport model dealing with particle-particle scatterings at high energies naturally tackles the interactions among different particles at high densities. Since the results of particle production given by transport models are frequently compared with nuclear experimental data from facilities worldwide, the scattering matrix element (determined by the strong interaction) among different particles at high densities in heavy-ion collisions are frequently adjusted. Therefore, hadronic transport model for relativistic heavy-ion collisions could give some certain properties of strongly interacting matter.

\begin{figure}[t]
\centering
\vspace{-0.25cm}
\includegraphics[width=0.5\textwidth]{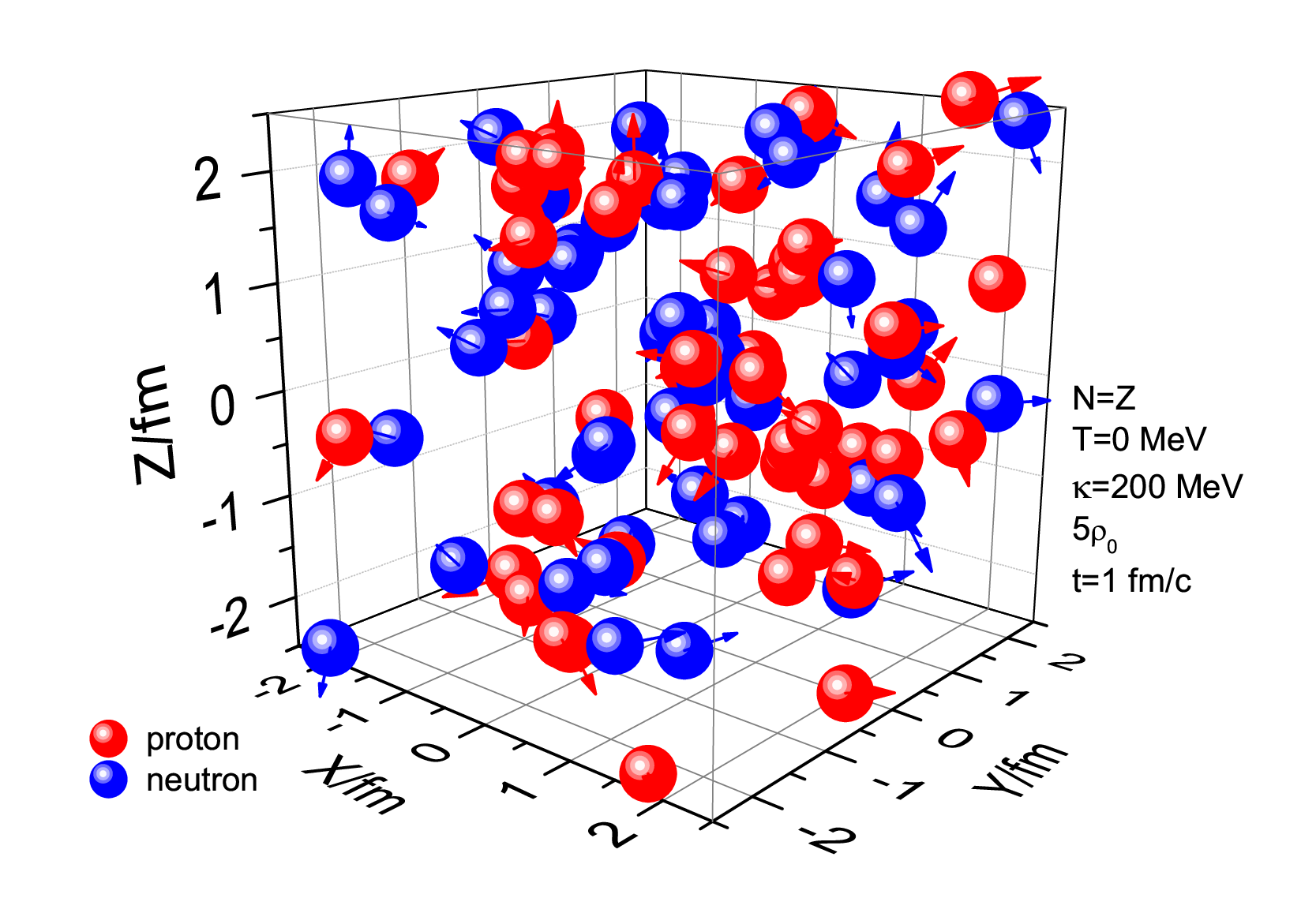}
\caption{Sketch of initial nuclear matter (using 1 run 1 test particle per nucleon as an example exhibition) at five times saturation density, i.e., 5$\rho_{0}$, in a box with periodic boundary condition. The dimensions of the cubic box are a = b = c = 5 fm. The position
of the center of box is (0, 0, 0). In practice, a particle that leaves the box on one side should enter it from the opposite side with the same momentum. In real simulations, 10 test-particle per nucleon and 1000 runs were employed. The time step is 0.2 fm/c and the code runs and stops at 1000 fm/c.} \label{3dbox}
\end{figure}
Figure~\ref{3dbox} shows nuclear matter in a box with periodic boundary condition, simulated by the pure hadronic transport model for relativistic heavy-ion collisions AMPT-HC. In the simulations, besides particle-particle collisions, mean-field potentials are added for most baryons and some mesons. The mean-field potential plays an important role in the long evolution process of dense matter. Coordinates of initial nucleons are randomly distributed in box while the initial momentum of each nucleon is decided by local Thomas-Fermi approximation with Fermi momentum calculated by its local density, i.e., $p_{F} = [3\pi^{2}\hbar^{3}\rho(r)_{n,p}]^{1/3}$. In this study, as the zero-temperature neutron stars are the most frequently discussed, the initial temperature of simulation is set to be zero.

\begin{figure}[t]
\centering
\vspace{-0.45cm}
\includegraphics[width=0.48\textwidth]{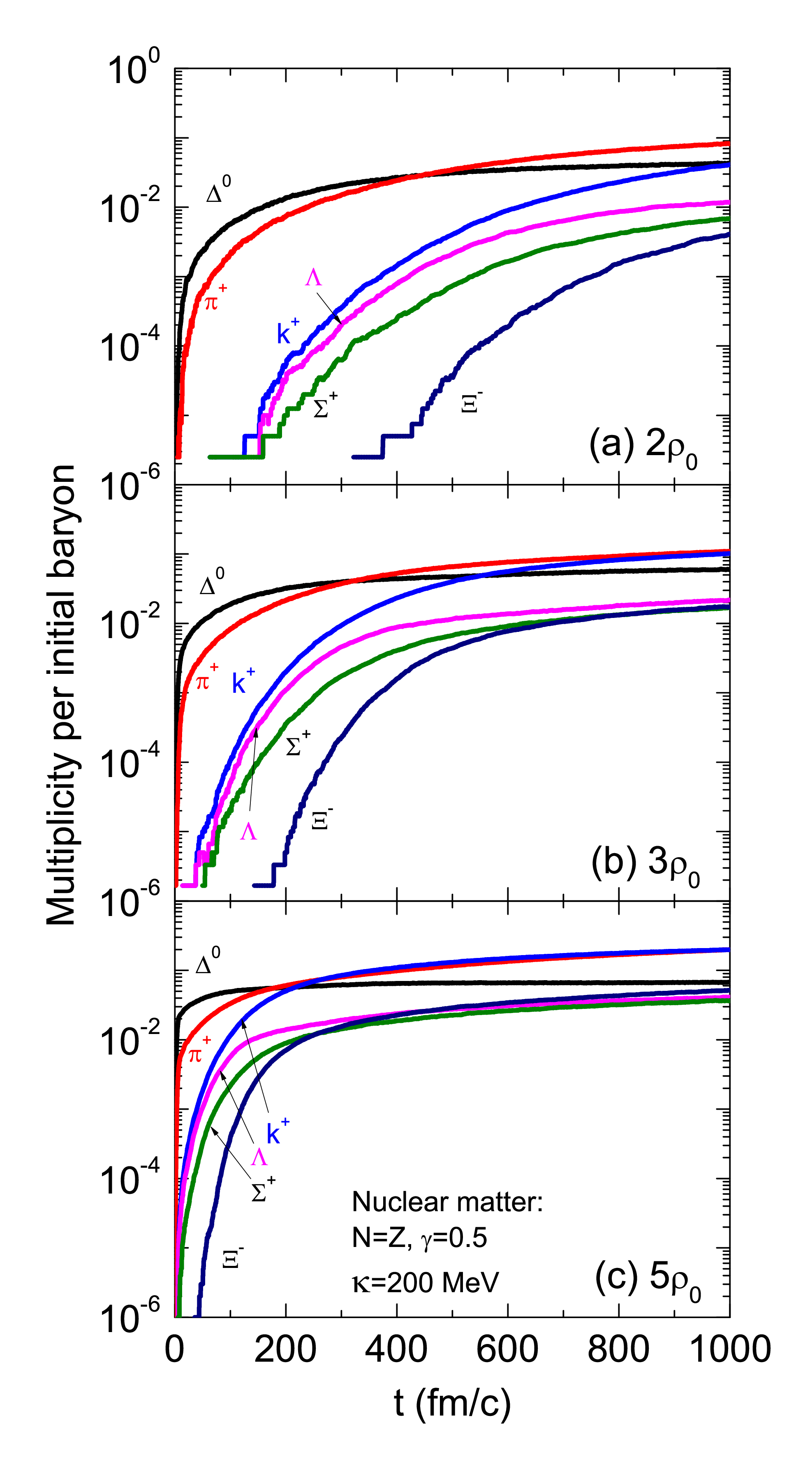}
\caption{Time evolution of several typical particle ($\Delta^{0}$, $\pi^{+}$, $K^{+}$, $\Lambda$, $\Sigma^{+}$ and $\Xi^{-}$) productions per initial baryon in dense matter with, respectively, 2, 3 and 5 times nuclear saturation density employing a pure hadronic transport model.} \label{evolu}
\end{figure}
To see how the non-nucleon particles are produced in dense matter, Figure~\ref{evolu} denoting time evolution of the non-nucleon particle productions in dense matter is plotted. It is seen that the non-nucleon particles are gradually produced as time increases. The non-strange particles $\Delta$ resonance and $\pi$ meson are first produced due to their low production thresholds. Their yields soon saturate as time increases, suggesting the balance of production and absorption is reached. As time increases, the non-strange particles experiencing multi-scatterings have enough energies to surpass threshold energies of strangeness productions. One thus sees that the singly strange particles $K^{+}$, $\Lambda$, $\Sigma^{+}$ are produced at later time and gradually reach saturation. The associated production of single strangeness is clearly demonstrated in Figure~\ref{evolu}. When there are enough singly strange particles in dense matter, the doubly strange $\Xi^{-}$ is ready to produce. It is clearly shown that the doubly strange $\Xi^{-}$ begins to produce at the latest time and gradually reaches saturation after 1000 fm/c. Overall, there is a rapid saturation on the production of non-nucleon particles in even denser matter. From Figure~\ref{evolu}, one sees that in the static dense matter, nucleons, resonances, mesons as well as single and double strangenesses all may exist. The existence of non-nucleon particles is supposed to alter the bulk properties of dense matter.

\begin{figure}[t]
\centering
\vspace{-0.45cm}
\includegraphics[width=0.48\textwidth]{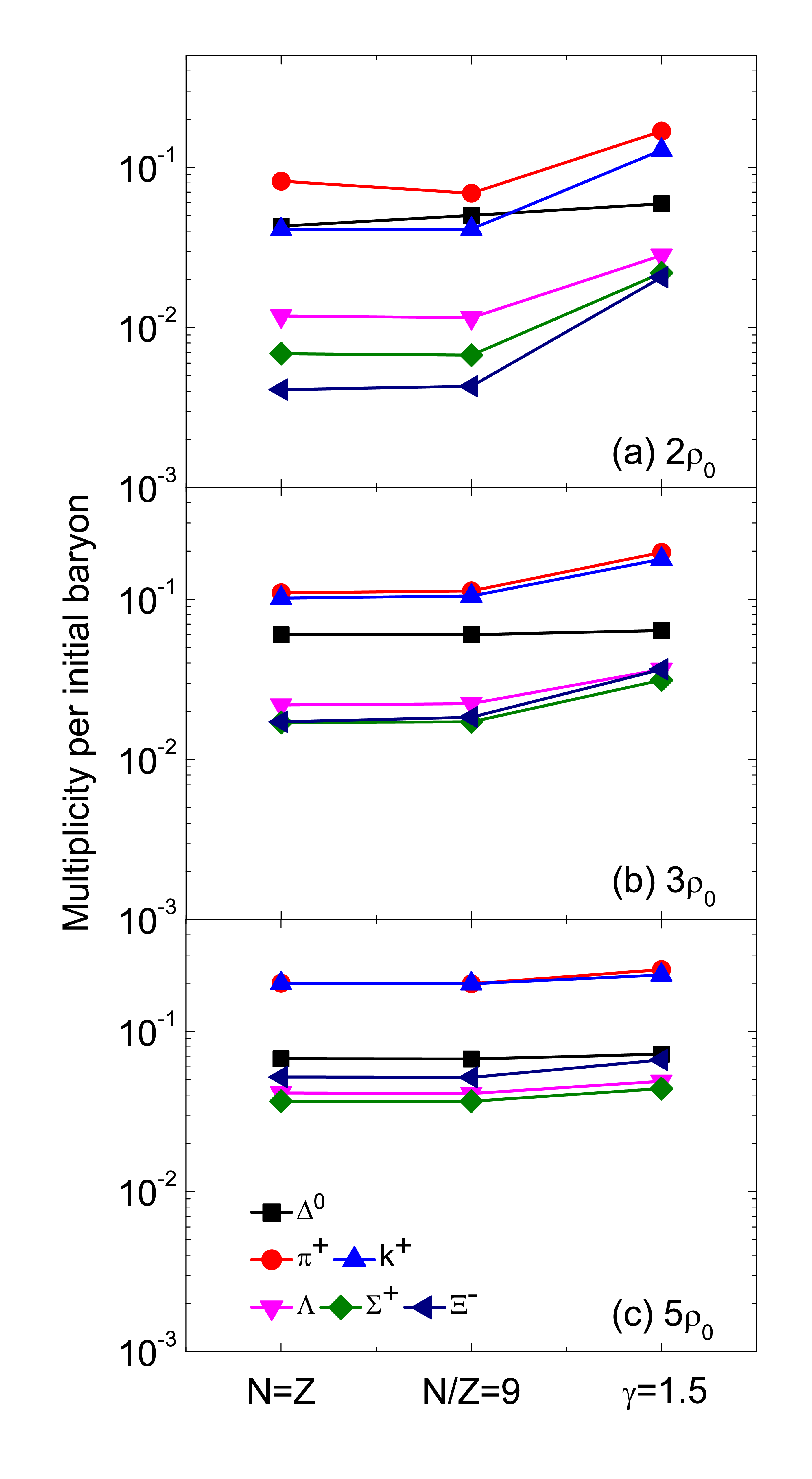}
\caption{Same as Figure~\ref{evolu}, but for final state situations with various asymmetries and symmetry energies. The default case for the symmetry energy is $E_{sym}(\rho/\rho_{0})=34.5(\rho/\rho_{0})^{\gamma = 0.5}$. The case ``$\gamma$ = 1.5'' denotes $E_{sym}(\rho/\rho_{0})=34.5(\rho/\rho_{0})^{\gamma = 1.5}$.} \label{final}
\end{figure}
In neutron star matter, the number of neutrons is generally considered to be about 9 times the number of protons. It is thus necessary to see if the previous result on the productions of non-nucleon particles changes in neutron star matter with super large asymmetry of neutron number and proton number. Also it is interesting to see if the symmetry energy plays a role in the production of non-nucleon particles in dense and asymmetric matter. Figure~\ref{final} shows the typical particle ($\Delta^{0}$, $\pi^{+}$, $K^{+}$, $\Lambda$, $\Sigma^{+}$ and $\Xi^{-}$) productions in dense symmetric (N = Z) and asymmetric (N/Z = 9) matter and the effects of EoS with the variety of symmetry energy are also shown. Firstly, it is seen that the asymmetry of dense matter in fact less affects the productions of non-nucleon particles. This is not only because the productions of these particles are less isospin-dependent but also most asymmetry effects have been smoothed out after many scatterings among particles. While the symmetry energy affects the productions of non-nucleon particles, especially at low densities. The stiff symmetry energy ($\gamma$ = 1.5) increases the average energy per nucleon, causes more energetic nucleon-nucleon collisions. Therefore more  non-nucleon particles are produced. Secondly, the fractions of $\pi^{+}$, $K^{+}$ mesons are overall larger than those of $\Delta^{0}$, $\Lambda$, $\Sigma^{+}$ and $\Xi^{-}$ except for twice saturation density with a soft EoS. The fractions of other particles could be estimated via isospin symmetry of related reaction channels. Finally, Figure~\ref{final} further demonstrates that even at two times saturation density, non-nucleon particles still occupy certain proportions. One thus should be careful when predicting the properties of dense matter above twice saturation density via nucleon-only many-body approaches.

\begin{figure}[t]
\centering
\includegraphics[width=0.45\textwidth]{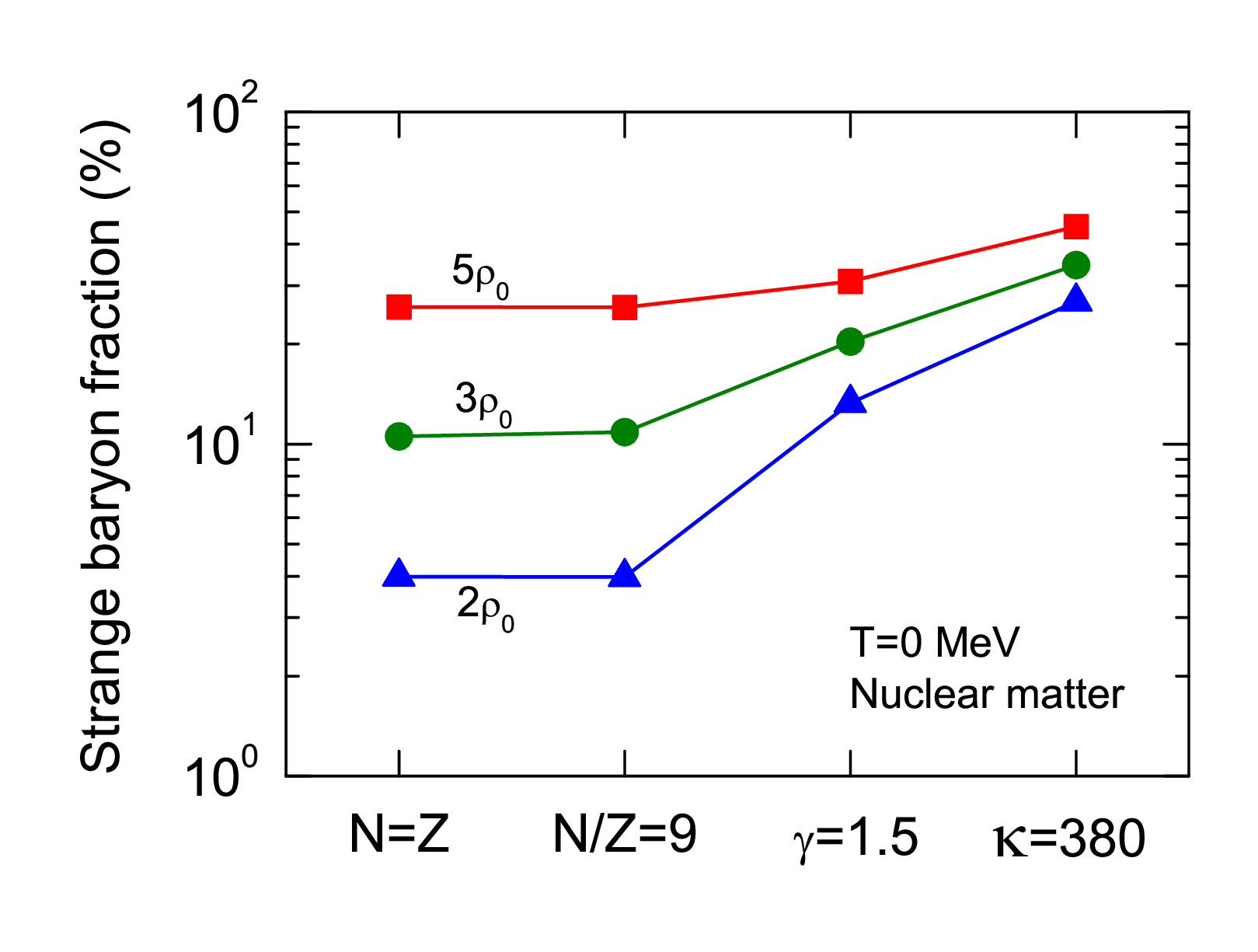}
\caption{Same as Figure~\ref{final}, but for the fraction of strange baryon in dense matter. The case ``$k$ = 380'' denotes a stiff EoS with an incompressibility coefficient $k$ = 380 MeV while the default value is $k$ = 200 MeV.} \label{ratio}
\end{figure}
The basic question of ``hyperon puzzle'' in the neutron star interior is the fraction of strangeness in neutron star matter. To show the fraction of strange baryon in neutron star matter, ratios of the strange baryon number over the total baryon number are demonstrated for symmetric and asymmetric matter with the variety of EoS. From Figure~\ref{ratio}, it is seen that the fractions of the strange baryons become larger for even denser matter. For dense matter at 2$\rho_{0}$, one can see that the fractions of strange baryons are about 4\% for symmetric (N = Z) and asymmetric (N/Z = 9) matter with a soft EoS. While such fractions increase evidently (roughly from 4\% to 10\% or 20\%) either for a stiff symmetry energy (``$\gamma$ = 1.5'') or further for a large incompressibility coefficient ($k$ = 380 MeV).
But the effects of the EoS on the fractions of strange baryons become less evident for even denser matter. For dense matter at 5$\rho_{0}$, the fractions of strange baryons could be as high as 25-50\%, depending on the EoS employed. Large fraction of the strange baryon in dense matter call for clear hyperon-nucleon and hyperon-hyperon interactions while studying the properties of neutron star matter via many-body calculations. It is gratifying to see that the study of hyperon-nucleon interaction is reignited at RHIC through the measurements of $\Lambda$ hyperon and hypernuclei directed flows \cite{star2023}.

The nuclear EoS affects the fraction of the strange baryon in dense, neutron star matter. While the strange composition in neutron star matter alters the stiffness of neutron star matter \cite{ch16}. The interplay of the EoS and the fraction of strangeness complicates the studies on the properties and the structure of dense, neutron star matter.

The present research on strangeness production in neutron star matter does not necessarily represent the actual situation inside neutron stars, but it provides new insights into the strange components in neutron stars.

%\section{Conclusions}
%
In summary, based on a relativistic transport model for heavy-ion collisions, the fractions of non-nucleon particles, especially for strange baryons in dense, neutron star matter are studied via box calculations. It is found that the fractions of strange baryons are not affected by the asymmetry of neutron number and proton number in dense matter, but the employed equation of state evidently affects the fractions of strange baryons. There exists a small fraction of strangeness in twice saturation density matter. For five times saturation density matter, the fraction of strange baryons can be as high as 25-50\%, depending on the equation of state used. Large fraction of the strange baryons in dense matter call for clear hyperon-nucleon and hyperon-hyperon interactions so as to determine the bulk properties of the neutron star matter.

%\section{Acknowledgments}
%
This work is supported by the National Natural Science Foundation of China under Grant Nos. 12275322, 12335008 and the Strategic Priority Research Program of Chinese Academy of Sciences with Grant No. XDB34030000.

\end{document}